\documentclass[aps,prb,twocolumn,longbibliography,showpacs]{revtex4-1}

\usepackage{graphicx}% Include figure files
\usepackage{dcolumn}% Align table columns on decimal point
\usepackage{bm}% bold math
\usepackage[version=3]{mhchem}
\usepackage{color}

\begin{document}

\title{First-principles theory of doping in layered oxide electrode materials}
\author{Khang Hoang}
\email{khang.hoang@ndsu.edu}
%\homepage[]{Your web page}
%\thanks{}
%\altaffiliation{}
\affiliation{Department of Physics \& Center for Computationally Assisted Science and Technology, North Dakota State University, Fargo, North Dakota 58108, USA.}

\date{\today}

\begin{abstract}

Doping lithium-ion battery electrode materials \ce{LiMO2} (M = Co, Ni, Mn) with impurities has been shown to be an effective way to optimize their electrochemical properties. Here, we report a detailed first-principles study of layered oxides \ce{LiCoO2}, \ce{LiNiO2}, and \ce{LiMnO2} lightly doped with transition-metal (Fe, Co, Ni, Mn) and non-transition-metal (Mg, Al) impurities using hybrid-density-functional defect calculations. We find that the lattice site preference is dependent on both the dopant's charge and spin states, which are coupled strongly to the local lattice environment and can be affected by the presence of co-dopant(s), and the relative abundance of the host compound's constituting elements in the synthesis environment. On the basis of the structure and energetics of the impurities and their complexes with intrinsic point defects, we determine all possible low-energy impurity-related defect complexes, thus providing defect models for further analyses of the materials. From a materials modeling perspective, these lightly doped compounds also serve as model systems for understanding the more complex, mixed-metal, \ce{LiMO2}-based battery cathode materials.  

\end{abstract}

% insert suggested PACS numbers in braces on next line
%\pacs{}
% insert suggested keywords - APS authors don't need to do this
%\keywords{}

%\maketitle must follow title, authors, abstract, \pacs, and \keywords
\maketitle

% body of paper here - Use proper section commands
% References should be done using the \cite, \ref, and \label commands

\section{Introduction}\label{sec;intro}

Layered transition-metal oxides \ce{LiMO2} (M = Co, Ni, Mn) and, especially, their derivatives such as Li(Ni,Co,Mn)O$_2$ [NCM, also known as NMC] and Li(Ni,Co,Al)O$_2$ [NCA] have been widely used as cathode materials in lithium-ion batteries.\cite{Whittingham_CR,AndreJMCA2015} These materials are known to exhibit rich defect physics resulted from the ability of the transition-metal ions to exist in different charge and spin states and the strong coupling between charge, spin, and local atomic structures.\cite{Hoang2014JMCA,Hoang2015PRA,Hoang2016CM} Doping \ce{LiMO2} with transition-metal and non-transition-metal impurities has been shown to be an effective way to optimize the electrochemical performance.\cite{Carewska1997,Tukamoto1997,Levasseur2002,Nobili2005EA,Kim2006JPS,Ceder1998,Luo2010JES,Zhou2009JES,Liang2014,Mukai2010JSSC,Stoyanova2010,Luo2012JMC,Bonda2008,Pouillerie2000} Here, the impurities (i.e., dopants) can be incorporated into \ce{LiMO2} at the transition-metal (M) and/or Li sites and the lattice site preference of some of the dopants may be dependent on the experimental conditions during synthesis. Understanding the effects of doping requires a detailed understanding of the interaction between the dopant and the host, including intrinsic point defects that may present in the host compound, under the synthesis conditions. 

Computationally, there have been a number of first-principles studies of doping in \ce{LiMO2} using density-functional theory (DFT) within the standard local-density approximation or generalized-gradient approximation or the DFT$+U$ extension (where $U$ is the on-site Hubbard correction).\cite{Koyama2014,Santana2014,Kim2012,Chen2014,Lee2017PRB,Prasad2005,Shukla2008,Kong2015JPCC} These studies have provided useful information on several aspects of the doped materials, including their atomic and electronic structure and the solubility of the dopants. However, the methods used in these previous studies are known to have limited predictive power in complex transition-metal oxides. Even within the DFT$+U$ extension, Santana et al.,\cite{Santana2014} for example, showed that the results are strongly dependent on the choice of the $U$ value for the $3d$ orbitals of the transition metal of the host. The problem becomes more challenging when the dopant itself is another transition metal. A more rigorous approach is thus needed to describe the physics of the doped \ce{LiMO2} systems, including the ability to properly address the coupling between charge, spin, and local atomic structures.

In this work, we carry out a detailed and systematic study of doping in \ce{LiMO2} using first-principles defect calculations based on a hybrid DFT/Hartree-Fock approach and our accumulated knowledge\cite{Hoang2014JMCA,Hoang2015PRA} of the bulk properties and intrinsic point defects in the layered oxides. Specific impurities considered include Mg and Al, which have been reported to have beneficial effects on the performance of \ce{LiMO2}, and transition metals Mn, Co, and Ni, often employed in ion substitution. From a materials modeling perspective, \ce{LiMO2} doped with a low concentration of impurities can be considered as model systems for understanding the more complex, mixed-metal oxides such as NCM and NCA. For example, as a first approximation, Ni-rich materials LiNi$_{1-x-y}$Co$_x$Mn$_y$O$_2$ and LiNi$_{1-x-y}$Co$_x$Al$_y$O$_2$,\cite{Myung2017ACSEL} can be regarded as \ce{LiNiO2} doped with (Co,Mn) and (Co,Al), respectively. Highly doped \ce{LiMO2} materials such as NCM$_{1/3}$ and NCA$_{1/3}$ (i.e., $x=y=1/3$) have been previously investigated.\cite{Hoang2016CM} The focus of the current work is on the lattice site preference of the dopants, charge and spin states of the dopants and the transition-metal ions in the host, and effects of co-doping. This study will provide physical insights into the dopant-host interaction and possible effects on the electrochemical performance of \ce{LiMO2}-based materials.

\section{Methodology}\label{sec;method}

\begin{figure*}
\centering
\includegraphics[width=18.0cm]{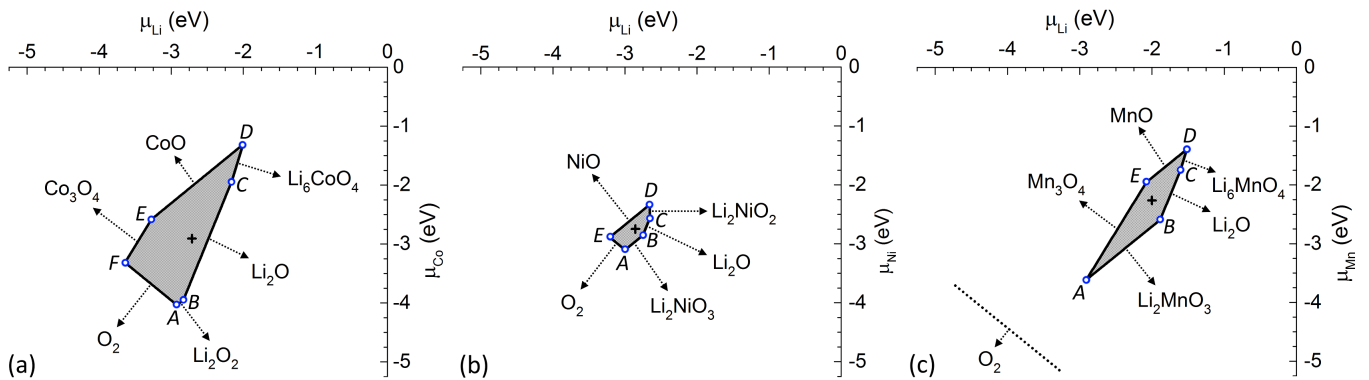}
\caption{Chemical-potential diagrams for \ce{LiMO2}: (a) M = Co, (b) M = Ni, and (c) M = Mn, produced using data from Refs.~\citenum{Hoang2014JMCA} and \citenum{Hoang2015PRA}. Only Li$-$M$-$O phases that define the stability region of \ce{LiMO2}, shown as a shaded polygon, are included; in (c), \ce{O2} is also included for reference. Point $X$ inside the stability region [not to be confused with defect X in Eq.~(\ref{eq:eform})] is marked by a cross. Explicit values of the atomic chemical potentials at representative points in the stability region are reported in Ref.~\citenum{SM}.}
\label{fig;chempot}
\end{figure*}

The total-energy calculations are based on DFT with the Heyd-Scuseria-Ernzerhof (HSE06) screened hybrid functional,\cite{heyd:8207} as implemented in the Vienna {\it Ab Initio} Simulation Package (\textsc{vasp}).\cite{VASP2} The Hartree-Fock mixing parameter ($\alpha$) and the screening length are set to the standard values of 0.25 and 10 {\AA}, respectively. The metal impurities in \ce{LiMO2}, in the dilute doping limit, are modeled using 108-atom hexagonal supercells\cite{Hoang2014JMCA,Hoang2015PRA} and a plane-wave basis-set cutoff of 500 eV; integrations over the supercell Brillouin zone are carried out using the $\Gamma$ point. In these calculations, the lattice parameters are fixed to the calculated bulk values of \ce{LiMO2} but all the internal coordinates are fully relaxed; the ferromagnetic spin configuration for the transition metal array in the lattice is used and spin polarization is included. Convergence with respect to self-consistent iterations is assumed when the total-energy difference between cycles is less than 10$^{-4}$ eV and the residual forces are less than 0.01 eV/{\AA}. We thus use the same calculation set-ups as in our previous work on bulk properties and intrinsic point defects in \ce{LiMO2}\cite{Hoang2014JMCA,Hoang2015PRA} to ensure the transferability of the results. 

The likelihood of an intrinsic defect, impurity or dopant (extrinsic defect), or defect complex X, hereafter often referred commonly to as ``defect'', in charge state $q$ being incorporated into a crystal is characterized by its formation energy, defined as
\begin{equation}\label{eq:eform}
\begin{split}
E^f({\mathrm{X}}^q)=&E_{\mathrm{tot}}({\mathrm{X}}^q)-E_{\mathrm{tot}}({\mathrm{bulk}})-\sum_{i}{n_i\mu_i} \\
                    &+q(E_{\mathrm{v}}+\mu_{e})+ \Delta^q ,
\end{split}
\end{equation}
where $E_{\mathrm{tot}}(\mathrm{X}^{q})$ and $E_{\mathrm{tot}}(\mathrm{bulk})$ are, respectively, the total energy of a supercell containing the defect X and that of a supercell of the perfect material. $\mu_{i}$ is the atomic chemical potential of species $i$ (and is referenced to bulk metals or O$_{2}$ molecules at 0 K). $n_{i}$ is the number of atoms of species $i$ that have been added ($n_{i}>$0) or removed ($n_{i}<$0) to form the defect. $\mu_{e}$ is the electronic chemical potential, i.e., the Fermi level, that is, as a convention, referenced to the valence-band maximum (VBM) in the perfect bulk ($E_{\mathrm{v}}$); the actual position of the Fermi level is determined by the charge neutrality condition that involves all defects and any other charge carriers that may be present in the material.\cite{Hoang2014JMCA,Hoang2015PRA} $\Delta^q$ is the correction term to align the electrostatic potentials of the bulk and defect supercells and to account for finite-size effects on the total energies of charged defects, estimated following the procedure of Freysoldt et al.\cite{Freysoldt2009} The total static dielectric constants used in the calculation of $\Delta^q$ are 13.02, 15.45, and 32.52 for \ce{LiCoO2}, \ce{LiNiO2}, and \ce{LiMnO2}, respectively.\cite{Hoang2014JMCA,Hoang2015PRA} 

The atomic chemical potentials of Li, M, and O in \ce{LiMO2} are subject to thermodynamic constraints and can be used to represent the experimental situations, e.g., during materials preparation. These constraints are to ensure that the host compound \ce{LiMO2} is thermodynamically stable.\cite{Hoang2014JMCA,Hoang2015PRA} Figure \ref{fig;chempot} shows the chemical-potential diagrams for \ce{LiMO2} in which the stability region is determined by considering equilibria with other Li$-$M$-$O phases. The results have been reported in Refs.~\citenum{Hoang2014JMCA} and \citenum{Hoang2015PRA} but are also produced here as we will frequently refer to these diagrams when discussing the results for the impurities. For the impurities in \ce{LiMO2}, the lower limit of their chemical potentials is minus infinity and the upper limit is zero, with respect to the total energy per atom of the bulk metals. Stronger bounds on the impurity chemical potentials can be estimated based on other solubility-limiting phases formed between the impurities and the host constituents.\cite{walle:3851} In the following, the chemical potentials of Mg, Al, Mn, Fe, Co, and Ni impurities are set as $\mu_{\rm Mg}$ = $-6.00$ eV, $\mu_{\rm Al}$ = $-9.00$ eV, $\mu_{\rm Mn}$ = $-6.00$ eV, $\mu_{\rm Fe}$ = $-5.00$ eV, $\mu_{\rm Co}$ = $-4.00$ eV, and $\mu_{\rm Ni}$ = $-4.00$ eV. These choices are somewhat arbitrary; however, they in no way affect the physics of what we are presenting as we are interested only in the relative formation energies of the impurities associated with different set of the atomic chemical potentials of the host constituents which correspond to different points in the chemical-potential diagram presented in Fig.~\ref{fig;chempot}. Formation energies for other values of the chemical potentials of the impurities, if desirable, can be easily obtained from the data we report.

We also investigate selected heavily doped \ce{LiMO2} systems, using smaller, 24-atom hexagonal supercells. In these calculations, one (or two, in the case of co-doping) atom of the host is substituted by the impurity atom(s), and the cell volume and shape and internal coordinates are all relaxed. Integrations over the Brillouin zone are carried out using a $\Gamma$-centered $11\times 6 \times 2$ $k$-point mesh to obtain high-quality electronic densities of states.

Finally, we note that smaller values of the mixing parameter $\alpha$ have also been employed in studies of the layered oxides using the HSE06 functional.\cite{Seo2015PRB} For reasonable choices of $\alpha$ values, however, the main difference is only in the calculated band-gap values; see Fig.~1 of Ref.~\citenum{SM}. As discussed in Ref.~\citenum{Hoang2014JMCA} and references therein, the defect formation energy at the Fermi level determined by the charge neutrality condition is usually not sensitive to the calculated band gap, provided that the calculations can capture the essential physics near the band edges.  

\section{Results and discussion}\label{sec;results}

\subsection{\ce{LiCoO2}}

\begin{figure}
\centering
%\vspace{-1cm}
%\hspace{-.5cm}
\includegraphics[width=8.6cm]{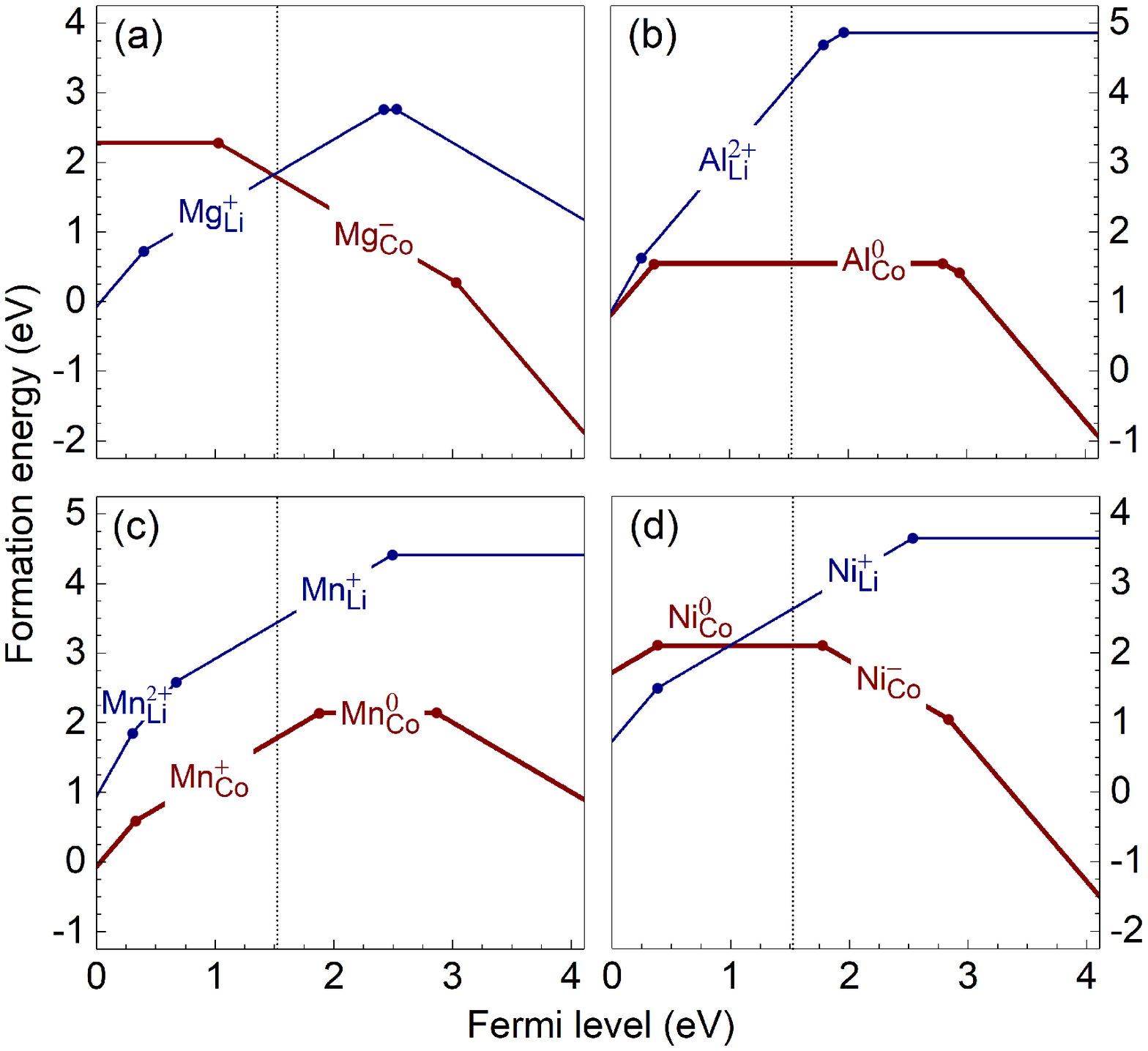}
%\vspace{0cm}
\caption{Formation energies of substitutional impurities at the Li and Co lattice sites in \ce{LiCoO2} obtained at point $X$ [marked by a cross in the chemical-potential diagram in Fig.~\ref{fig;chempot}(a)], plotted as a function of Fermi level from the VBM to the conduction-band minimum (CBM) of the undoped compound: (a) Mg, (b) Al, (c) Mn, and (d) Ni. The slope in the energy plots indicates the charge state ($q$). For each defect, only the true charge states are indicated. The vertical dotted line marks the Fermi level of undoped \ce{LiCoO2}, $\mu_{e}^{\rm{int}}$, determined by the intrinsic point defects as reported in Ref.~\citenum{Hoang2014JMCA}}
\label{fig;lco}
\end{figure}

Figure~\ref{fig;lco} shows the formation energies of substitutional Mg, Al, Mn, Fe, and Ni impurities at the Co and Li sites in \ce{LiCoO2}, obtained under the conditions at point $X$ in the chemical-potential diagram [Fig.~\ref{fig;chempot}(a)]. We find that each impurity has only one or two true charge states (hereafter also called elementary defects) among possible values of $q$; the other charge states correspond to complexes consisting of the elementary defects and hole ($\eta^+$) or electron ($\eta^-$) polaron(s). Note that, in \ce{LiCoO2}, Co is stable as low-spin Co$^{3+}$; $\eta^+$ ($\eta^-$) is the localized hole (electron) and local lattice distortion associated with the low-spin Co$^{4+}$ (high-spin Co$^{2+}$) ion at the Co lattice site.\cite{Hoang2014JMCA} Taken the Mn impurity as an example, Mn$_{\rm Co}^0$ (i.e., high-spin Mn$^{3+}$ at the Co site) and Mn$_{\rm Co}^+$ (i.e., Mn$^{4+}$ at the Co site) are elementary defects, as indicated in Fig.~\ref{fig;lco}(c), whereas Mn$_{\rm Co}^-$ is a complex of Mn$_{\rm Co}^0$ and $\eta^-$ and Mn$_{\rm Co}^{2+}$ is a complex of Mn$_{\rm Co}^+$ and $\eta^+$. In this case, $q = 0$ and $+$ are true charge states; $q = -$ and $2+$ are regarded only as nominal charge states. Similarly, on the Li sublattice, Mn$_{\rm Li}^+$ (i.e., high-spin Mn$^{2+}$ at the Li site) and Mn$_{\rm Li}^{2+}$ (i.e., high-spin Mn$^{3+}$ at the Li site) are elementary defects; Mn$_{\rm Li}^0$ is a complex of Mn$_{\rm Li}^+$ and $\eta^-$. 

In the following, we focus on the structure and energetics of the most stable defect configurations at the Fermi-level position of undoped \ce{LiCoO2}, specifically at $\mu_{e}^{\rm{int}}$ that is determined by the intrinsic point defects in the undoped compound for a given set of the atomic chemical potentials,\cite{Hoang2014JMCA} e.g., as indicated by the vertical dotted line in Fig.~\ref{fig;lco}. Under the conditions at point $X$ in Fig.~\ref{fig;chempot}(a), chosen as representative conditions for the presentation purpose, Mg is most stable as Mg$_{\rm Co}^-$ (Mg$_{\rm Li}^+$) on the Co (Li) sublattice, whereas Al is most stable as Al$_{\rm Co}^0$ (Al$_{\rm Li}^{2+}$), Mn as Mn$_{\rm Co}^+$ (Mn$_{\rm Li}^+$), and Ni as Ni$_{\rm Co}^0$ (Ni$_{\rm Li}^+$) on the Co (Li) sublattice; see Fig.~\ref{fig;lco}. Fe (not included in the figure) is most stable as Fe$_{\rm Co}^0$ (Fe$_{\rm Li}^+$) on the Co (Li) sublattice. It is noted that, since the intrinsic-defect landscape (and hence the Fermi level $\mu_{e}^{\rm{int}}$) varies as a function of the atomic chemical potentials,\cite{Hoang2014JMCA} the lattice site and charge (and spin) state preference of the impurities may also be different for different points in the chemical-potential diagram in Fig.~\ref{fig;chempot}(a), as discussed in more detail below. 

\begin{figure}
\centering
%\vspace{-1cm}
%\hspace{-.5cm}
\includegraphics[width=8.6cm]{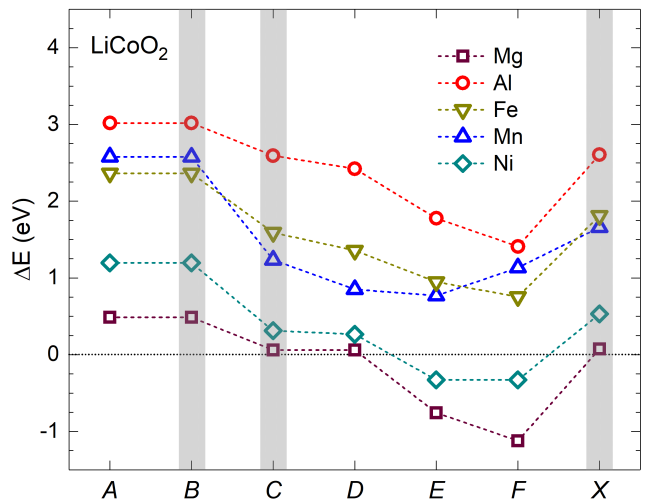}
%\vspace{0cm}
\caption{Difference between the formation energies at the Co and Li sites, obtained under the conditions at points $A-F$ and $X$ in Fig.~\ref{fig;chempot}(a). $\Delta E > 0$ means the impurity is energetically more favorable at the Co site. Points $B$, $C$, and $X$ can be regarded as representing more realistic synthesis conditions.}
\label{fig;lco;diff}
\end{figure}

\begin{table*}
\caption{\ Defect models for the impurities (dopants) in \ce{LiMO2} (M = Co, Ni, Mn) under the conditions at different points in the chemical-potential diagrams (Fig.~\ref{fig;chempot}). Only the most stable configurations are included; other configurations that are close in energy are listed in the footnotes.}\label{tab;models}
\begin{center}
\begin{ruledtabular}
\begin{tabular}{lllllllll}
%\hline
&Dopant&$A$&$B$&$C$&$D$&$E$&$F$&$X$ \\
\hline
\ce{LiCoO2}
&Mg&Mg$_{\rm Co}^-$$-$$\eta^+$&Mg$_{\rm Co}^-$$-$$\eta^+$&Mg$_{\rm Co}^-$$-$$\eta^+$$^a$&Mg$_{\rm Co}^-$$-$Co$_{\rm Li}^+$$^b$&Mg$_{\rm Li}^+$$-$$V_{\rm Li}^-$&Mg$_{\rm Li}^+$$-$$V_{\rm Li}^-$&Mg$_{\rm Co}^-$$-$$\eta^+$$^c$ \\
&Al&Al$_{\rm Co}^0$&Al$_{\rm Co}^0$&Al$_{\rm Co}^0$&Al$_{\rm Co}^0$&Al$_{\rm Co}^0$&Al$_{\rm Co}^0$&Al$_{\rm Co}^0$ \\
&Fe&Fe$_{\rm Co}^0$&Fe$_{\rm Co}^0$&Fe$_{\rm Co}^0$&Fe$_{\rm Co}^0$&Fe$_{\rm Co}^0$&Fe$_{\rm Co}^0$&Fe$_{\rm Co}^0$ \\
&Mn&Mn$_{\rm Co}^+$$-$Li$_{\rm Co}^{2-}$$-$$\eta^+$&Mn$_{\rm Co}^+$$-$Li$_{\rm Co}^{2-}$$-$$\eta^+$&Mn$_{\rm Co}^0$&Mn$_{\rm Co}^0$&Mn$_{\rm Co}^+$$-$$V_{\rm Li}^-$$^d$&Mn$_{\rm Co}^+$$-$$V_{\rm Li}^-$&Mn$_{\rm Co}^0$ \\
&Ni&Ni$_{\rm Co}^0$&Ni$_{\rm Co}^0$&Ni$_{\rm Co}^0$&Ni$_{\rm Co}^0$&Ni$_{\rm Li}^+$$-$$V_{\rm Li}^-$&Ni$_{\rm Li}^+$$-$$V_{\rm Li}^-$&Ni$_{\rm Co}^0$ \\
\ce{LiNiO2}
&Mg&Mg$_{\rm Ni}^-$$-$$\eta^+$&Mg$_{\rm Ni}^-$$-$$\eta^+$&Mg$_{\rm Ni}^-$$-$$\eta^+$$^e$&Mg$_{\rm Li}^+$$-$$\eta^-$$^f$&Mg$_{\rm Li}^+$$-$$\eta^-$$^f$&&Mg$_{\rm Ni}^-$$-$$\eta^+$$^g$ \\
&Al&Al$_{\rm Ni}^0$&Al$_{\rm Ni}^0$&Al$_{\rm Ni}^0$&Al$_{\rm Ni}^0$&Al$_{\rm Ni}^0$&&Al$_{\rm Ni}^0$ \\
&Fe&Fe$_{\rm Ni}^0$&Fe$_{\rm Ni}^0$&Fe$_{\rm Ni}^0$&Fe$_{\rm Ni}^0$&Fe$_{\rm Ni}^0$&&Fe$_{\rm Ni}^0$ \\
&Mn&Mn$_{\rm Ni}^+$$-$$\eta^-$$^h$&Mn$_{\rm Ni}^+$$-$$\eta^-$$^h$&Mn$_{\rm Ni}^+$$-$$\eta^-$&Mn$_{\rm Ni}^+$$-$$\eta^-$&Mn$_{\rm Ni}^+$$-$$\eta^-$&&Mn$_{\rm Ni}^+$$-$$\eta^-$ \\
&Co&Co$_{\rm Ni}^0$&Co$_{\rm Ni}^0$&Co$_{\rm Ni}^0$&Co$_{\rm Ni}^0$&Co$_{\rm Ni}^0$&&Co$_{\rm Ni}^0$ \\
\ce{LiMnO2}
&Mg&Mg$_{\rm Li}^+$$-$$V_{\rm Li}^-$$^i$&Mg$_{\rm Mn}^-$$-$$\eta^+$$^j$&Mg$_{\rm Mn}^-$$-$Mn$_{\rm Li}^+$&Mg$_{\rm Mn}^-$$-$Mn$_{\rm Li}^+$&Mg$_{\rm Mn}^-$$-$Mn$_{\rm Li}^+$&&Mg$_{\rm Mn}^-$$-$Mn$_{\rm Li}^+$$^k$ \\
&Al&Al$_{\rm Mn}^0$&Al$_{\rm Mn}^0$&Al$_{\rm Mn}^0$&Al$_{\rm Mn}^0$&Al$_{\rm Mn}^0$&&Al$_{\rm Mn}^0$ \\
&Fe&Fe$_{\rm Mn}^0$&Fe$_{\rm Mn}^0$&Fe$_{\rm Mn}^0$&Fe$_{\rm Mn}^0$&Fe$_{\rm Mn}^0$&&Fe$_{\rm Mn}^0$ \\
&Co&Co$_{\rm Mn}^0$$^l$&Co$_{\rm Mn}^0$&Co$_{\rm Mn}^-$$-$Mn$_{\rm Li}^+$$^m$&Co$_{\rm Mn}^-$$-$Mn$_{\rm Li}^+$&Co$_{\rm Mn}^-$$-$Mn$_{\rm Li}^+$&&Co$_{\rm Mn}^-$$-$Mn$_{\rm Li}^+$$^n$ \\
&Ni&Ni$_{\rm Mn}^-$$-$$\eta^+$$^o$&Ni$_{\rm Mn}^-$$-$$\eta^+$$^p$&Ni$_{\rm Mn}^-$$-$Mn$_{\rm Li}^+$$^q$&Ni$_{\rm Mn}^-$$-$Mn$_{\rm Li}^+$&Ni$_{\rm Mn}^-$$-$Mn$_{\rm Li}^+$&&Ni$_{\rm Mn}^-$$-$Mn$_{\rm Li}^+$$^r$ \\
\end{tabular}
\end{ruledtabular}
\end{center}
\begin{flushleft}
{\small
$^a$Mg$_{\rm Co}^-$$-$Co$_{\rm Li}^+$ ($+0.01$ eV) and Mg$_{\rm Li}^+$$-$$\eta^-$ ($+0.05$ eV).
$^b$Mg$_{\rm Li}^+$$-$$\eta^-$ ($+0.05$ eV).
$^c$Mg$_{\rm Li}^+$$-$$V_{\rm Li}^-$ ($+0.21$ eV).
$^d$Mn$_{\rm Co}^0$ ($+0.19$ eV).
$^e$Mg$_{\rm Li}^+$$-$$\eta^-$ ($+0.15$ eV).
$^f$Mg$_{\rm Ni}^-$$-$$\eta^+$ ($+0.08$ eV) and Mg$_{\rm Ni}^-$$-$Ni$_{\rm Li}^+$ ($+0.20$ eV).
$^g$Mg$_{\rm Li}^+$$-$$\eta^-$ ($+0.14$ eV).
$^h$Mn$_{\rm Ni}^+$$-$Li$_{\rm Ni}^{2-}$$-$$\eta^+$ ($+0.23$ eV).
$^i$Mg$_{\rm Mn}^-$$-$$\eta^+$ ($+0.12$ eV).
$^j$Mg$_{\rm Li}^+$$-$Li$_{\rm Mn}^{2-}$$-$$\eta^+$ ($+0.14$ eV) and Mg$_{\rm Mn}^-$$-$Mn$_{\rm Li}^+$ ($+0.23$ eV).
$^k$Mg$_{\rm Mn}^-$$-$$\eta^+$ ($+0.21$ eV).
$^l$Co$_{\rm Li}^+$$-$$V_{\rm Li}^-$ ($+0.21$ eV).
$^m$Co$_{\rm Mn}^0$ ($+0.15$ eV) and Co$_{\rm Mn}^-$$-$Li$_{i}^+$ ($+0.21$ eV).
$^n$Co$_{\rm Mn}^0$ ($+0.04$ eV).
$^o$Ni$_{\rm Li}^+$$-$$V_{\rm Li}^-$ ($+0.10$ eV) and Ni$_{\rm Mn}^-$$-$Mn$_{\rm Li}^+$ ($+0.21$ eV).
$^p$Ni$_{\rm Mn}^-$$-$Li$_{i}^+$ ($+0.15$ eV) and Ni$_{\rm Mn}^-$$-$Mn$_{\rm Li}^+$ ($+0.21$ eV).
$^q$Ni$_{\rm Mn}^-$$-$Li$_{i}^+$ ($+0.22$ eV).
$^r$Ni$_{\rm Mn}^-$$-$$\eta^+$ ($+0.24$ eV).
}
\end{flushleft}
\end{table*}

To quantify the lattice site preference of the impurities over the substitutional sites in \ce{LiMO2}, we define the energy difference
\begin{equation}
\Delta E = E^f({\mathrm{X}}^{q1}_{\rm Li}) - E^f({\mathrm{X}}^{q2}_{\rm M}),
\end{equation}
where $E^f({\mathrm{X}}^{q1}_{\rm Li})$ and $E^f({\mathrm{X}}^{q2}_{\rm M})$ are the formation energies (at $\mu_{e}^{\rm{int}}$) of the lowest-energy defect configurations at the Li and M sites, respectively. Here, $\Delta E > 0$ means the impurity X is energetically more favorable as X$_{\rm M}^{q2}$ (i.e., at the M site) than as X$_{\rm Li}^{q1}$ (the Li site), whereas $\Delta E \sim 0$ indicates that the impurity can be incorporated both on the M and Li sites with almost equal concentrations.

Figure~\ref{fig;lco;diff} shows the formation-energy difference between the Co and Li sites under the conditions at different points in the chemical-potential diagram [Fig.~\ref{fig;chempot}(a)]. We find that the impurities prefer the Co site over the Li site ($\Delta E >0$), except Mg and Ni which can be energetically more favorable at the Co or Li site depending on the synthesis conditions. More specifically, when considered as isolated defects, Al is most stable as Al$_{\rm Co}^0$ (i.e., Al$^{3+}$ at the Co site), Fe as Fe$_{\rm Co}^0$ (i.e., high-spin Fe$^{3+}$ at the Co site), and Mn as Mn$_{\rm Co}^+$, independent of the atomic chemical potentials. Mg is most stable as Mg$_{\rm Co}^-$ (i.e., Mg$^{2+}$ at the Co site) under the conditions at points $A-D$ and $X$ or Mg$_{\rm Li}^+$ (i.e., Mg$^{2+}$ at the Li site) at points $E$ and $F$. Finally, Ni is most stable as Ni$_{\rm Co}^0$ (i.e., low-spin Ni$^{3+}$ at the Co site) under the conditions at points $A-C$ and $X$, Ni$_{\rm Co}^-$ (i.e., high-spin Ni$^{2+}$ at the Co site) at point $D$, or Ni$_{\rm Li}^+$ (i.e., Ni$^{2+}$ at the Li site) at points $E$ and $F$. 

Overall, the lattice site preference of the impurities in \ce{LiCoO2} does not have a simple dependence on the ionic-radius difference between the dopant and the substituted host ion, but is determined by both the dopant's charge and spin states and the relative abundance of the host's constituting elements in the synthesis environment. We note that the charge and spin states are coupled strongly to the local lattice environment and thus also determine the dopant's ionic radius. The relative abundance of the host constituents is represented by the atomic chemical potentials in our computational approach (see Sec.~\ref{sec;method}).

\begin{figure*}
\centering
\includegraphics[width=18cm]{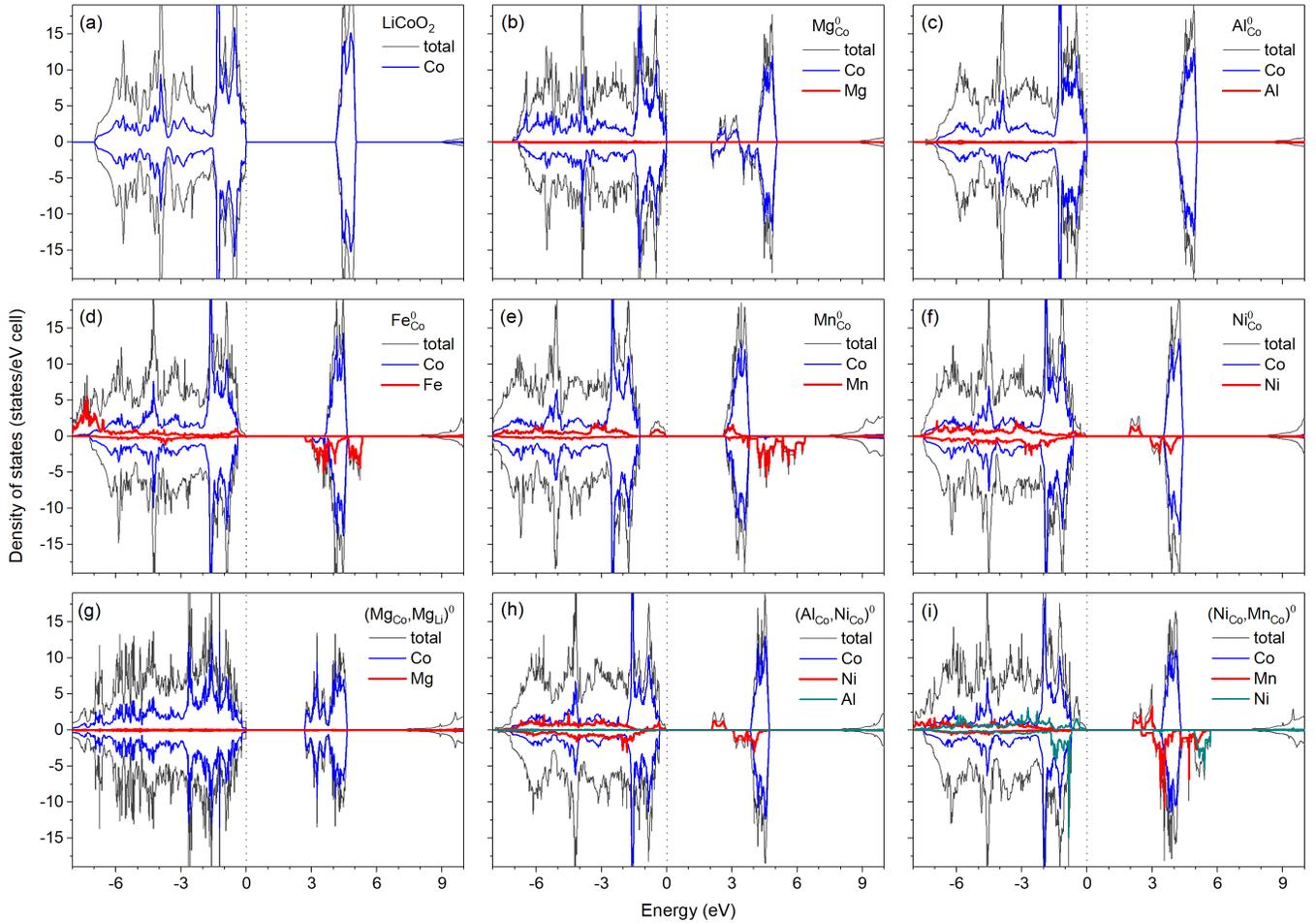}
\caption{Total and atomic-projected electronic densities of states of (a) undoped \ce{LiCoO2} and heavily doped \ce{LiCoO2} systems containing (b) Mg$_{\rm Co}^0$, (c) Al$_{\rm Co}^0$, (d) Fe$_{\rm Co}^0$, (e) Mn$_{\rm Co}^0$, (f) Ni$_{\rm Co}^0$, (g) (Mg$_{\rm Co}$,Mg$_{\rm Li}$)$^0$, (h) (Al$_{\rm Co}$,Ni$_{\rm Co}$)$^0$, and (i) (Ni$_{\rm Co}$,Mn$_{\rm Co}$)$^0$. The zero of energy is set to the highest occupied state.}
\label{fig;dos}
\end{figure*}

We also consider a neutral (Ni,Mn) pair in \ce{LiCoO2} with both dopants on the Co sublattice and find that it is most stable as (Ni$^{2+}$,Mn$^{4+}$), i.e., a complex of Ni$_{\rm Co}^-$ and Mn$_{\rm Co}^+$, which indicates charge transfer between the two impurities. The lowest-energy configuration of the pair corresponds to the shortest distance (2.82 {\AA}) between oppositely charged defects Mn$_{\rm Co}^+$ and Ni$_{\rm Co}^-$, as expected due to the Coulomb interaction. In this configuration, the binding energy ($E_b$) of the complex is 0.67 eV with respect to isolated Ni$_{\rm Co}^-$ and Mn$_{\rm Co}^+$. It is noted that, as an isolated defect, Ni can be stable as Ni$^{2+}$ or Ni$^{3+}$ on the the Co sublattice, as presented earlier. Yet in the (Ni,Mn) pair the Ni$^{2+}$ is always more stable. The results thus indicate that the dopant's charge and spin states can be affected by the presence of a co-dopant. It is also this impurity-impurity interaction between the transition-metal ions on the Co sublattice that causes the charge ordering observed in NCM$_{1/3}$.\cite{Hoang2016CM} For comparison, we find that a neutral (Ni,Al) pair in \ce{LiCoO2} is stable as (Ni$^{3+}$,Al$^{3+}$), i.e., Ni$_{\rm Co}^0$ and Al$_{\rm Co}^0$, and its total energy is almost independent of the pair distance ($E_b = 0$).

The impurities may occur in the material not as isolated defects but complexes with the intrinsic defects. On the basis of our results above regarding the most stable configurations of the impurities as well as those for the intrinsic defects under different sets of the atomic chemical potentials reported in Ref.~\citenum{Hoang2014JMCA}, we carry out calculations for all possible low-energy impurity-related neutral complexes and identify the energetically most stable configurations. Table~\ref{tab;models} summarizes the lowest-energy impurity-related defect complexes in \ce{LiCoO2} under various preparation conditions. Since \ce{LiCoO2} is often prepared under Li-rich conditions,\cite{Hoang2014JMCA,Chernova2011} the typical experimental conditions can be identified with approximately the region enclosing points $B$, $C$, and $X$ in Fig.~\ref{fig;chempot}(a). Under these conditions, Mg can be present in \ce{LiCoO2} in the form of the neutral complex Mg$_{\rm Co}^-$$-$$\eta^+$; i.e., the incorporation of the Mg impurity at the Co site is charge-compensated by the creation of $\eta^+$ (i.e., Co$^{4+}$) in the host. The complex has a binding energy $E_b = 0.65$ eV with respect to its isolated constituents. Al, Fe, and Ni are, on the other hand, most stable as Al$_{\rm Co}^0$, Fe$_{\rm Co}^0$, and Ni$_{\rm Co}^0$, respectively. Finally, Mn can be present in the form of the neutral complex Mn$_{\rm Co}^+$$-$Li$_{\rm Co}^{2-}$$-$$\eta^+$ (at point $B$; $E_b = 2.19$ eV) or Mn$_{\rm Co}^0$ (points $C$ and $X$). Other defect complexes listed in Table~\ref{tab;models} include Mg$_{\rm Co}^-$$-$Co$_{\rm Li}^+$ ($E_b = 0.49$ eV), Mg$_{\rm Li}^+$$-$$\eta^-$ ($E_b = 0.50$ eV), Mg$_{\rm Li}^+$$-$$V_{\rm Li}^-$ ($E_b = 0.51$ eV), Mn$_{\rm Co}^+$$-$$V_{\rm Li}^-$ ($E_b = 0.42$ eV), and Ni$_{\rm Li}^+$$-$$V_{\rm Li}^-$ ($E_b = 0.48$ eV). 

In the case where the formation-energy difference $\Delta E$ is small, one should expect that the dopant is incorporated at both the Co and Li sites. For example, Mg-doped \ce{LiCoO2} prepared under the conditions at point $C$ in Fig.~\ref{fig;chempot}(a) has two defect complexes with almost equal formation energies: Mg$_{\rm Co}^-$$-$$\eta^+$ and Mg$_{\rm Li}^+$$-$$\eta^-$, see Table~\ref{tab;models}. These two complexes can combine to form Mg$_{\rm Co}^-$$-$Mg$_{\rm Li}^+$ which has Mg over both the Co and Li sites. Besides, it is expected that both isolated impurities and impurity-related defect complexes are present in real samples of the doped \ce{LiCoO2} materials. The relative concentration of the complexes versus their isolated constituents is likely to be dependent on the total concentration of the dopants (and co-dopants, if present) and/or intrinsic point defects and their distribution in the materials.

Figure \ref{fig;dos} shows the electronic densities of states (DOS) of selected heavily doped \ce{LiCoO2} systems, obtained in calculations using the 24-atom supercells (see Sec.~\ref{sec;method}). The defect models chosen for these calculations are those with the lowest energies under the conditions at points $C$ and $X$ reported in Table~\ref{tab;models}; i.e., Mg$_{\rm Co}^0$ (i.e., Mg$_{\rm Co}^-$$-$$\eta^+$), Al$_{\rm Co}^0$, Fe$_{\rm Co}^0$, Mn$_{\rm Co}^0$, and Ni$_{\rm Co}^0$. The DOS of \ce{LiCoO2} containing (Mg$_{\rm Co}$,Mg$_{\rm Li}$)$^0$ (i.e., Mg$_{\rm Co}^-$ and Mg$_{\rm Li}^+$), (Al$_{\rm Co}$,Ni$_{\rm Co}$)$^0$ (i.e., Al$_{\rm Co}^0$ and Ni$_{\rm Co}^0$), or (Ni$_{\rm Co}$,Mn$_{\rm Co}$)$^0$ (i.e., Ni$_{\rm Co}^-$ and Mn$_{\rm Co}^+$) is also included. We find that the material stays non-metallic upon doping, which is consistent with our analysis of defect physics in \ce{LiCoO2} reported previously according to which (charged) impurities are charge-compensated by intrinsic point defects and the Fermi level of the system cannot be shifted to the VBM or CBM.\cite{Hoang2014JMCA} Focusing on the electronic structure near the band edges as it is relevant to the electrochemical properties,\cite{Hoang2015PRA} we find that, compared to the perfect bulk [Fig.~\ref{fig;dos}(a)], the Mg doping strongly disturbs the conduction-band bottom with additional electronic states at $\sim$2.0$-$4.0 eV coming from the Co$^{4+}$ ion (i.e., $\eta^+$), see Fig.~\ref{fig;dos}(b). The doping of \ce{LiCoO2} with Al has almost no change in the electronic structure near the band edges, see Fig.~\ref{fig;dos}(c), whereas the Fe doping disturbs the band edges of the host compound, see Fig.~\ref{fig;dos}(d). The Mn doping introduces electronic states associated with Mn$^{3+}$ at $\sim$$-1.0$$-$0.0 eV, see Fig.~\ref{fig;dos}(e), whereas the Ni doping introduces electronic states associated with Ni$^{3+}$ at $\sim$2.0$-$3.5 eV, see Fig.~\ref{fig;dos}(f). The electronic structure of the (Al,Ni)-doped system reflects that of the Al- and Ni-doped systems, see Fig.~\ref{fig;dos}(h). The electronic structure of the (Ni,Mn)-doped system is characterized by the electronic states associated with Ni$^{2+}$ at the valence-band top and those of Mn$^{4+}$ at the conduction-band bottom, see Fig.~\ref{fig;dos}(i). The calculated electronic structure is thus consistent with the details of the defect models.

Experimentally, it has been widely reported that Mg doping in \ce{LiCoO2} enhances the electronic conductivity.\cite{Carewska1997,Tukamoto1997,Levasseur2002,Nobili2005EA,Kim2006JPS} Levasseur et al.,\cite{Levasseur2002} for example, found that the activation energy for electronic conduction decreases as increasing the Mg concentration. This observation can be understood in terms of the results presented above in which the incorporation of Mg into the material at the Co site results in the formation of the Mg$_{\rm Co}^-$ defect and the hole polaron $\eta^+$. Equivalently, the incorporation of Mg into the Co sublattice can be regarded as acceptor-like doping:\cite{Hoang2012JPS} Once the concentration of Mg$_{\rm Co}^-$ is higher than that of the lowest-energy negatively charged intrinsic defect, this acceptor-like defect will shift the Fermi level of the system from the position $\mu_{e}^{\rm{int}}$ of undoped \ce{LiCoO2} slightly toward to the VBM, thus lowering the formation energy and hence increasing the concentration of $\eta^+$ (and other positively charged defects). The polarons remain in the samples after preparation and act as athermal, preexisting current-carrying defects during subsequent electrical conductivity measurements or facilitate lithium extraction at the beginning of the delithiation process.\cite{Hoang2014JMCA,Hoang2017PRM} Regarding the lattice site preference of Mg, Shim et al.\cite{Shim2014CM} observed that the ratio between the Mg levels at the Co and Li sites is dependent on the thermal treatment temperature, which is consistent with our analysis regarding the dependence on the chemical potentials.

Regarding the other dopants, Al, Fe, and Mn were reported to be incorporated into \ce{LiCoO2} at the Co site,\cite{Myung2001SSI,Menetrier2005CM,Luo2012JMC} in agreement with our results. Luo et al.\cite{Luo2012JMC} was able to stabilize Mn$^{4+}$ in the material. The defect model in this case could be one of those Mn$_{\rm Co}^+$-containing complexes listed in Table~\ref{tab;models}. As for Ni, Liang et al.\cite{Liang2014} found that the amount of Ni that goes into the Li sublattice can be controlled through tuning the Li content, which is equivalent to tuning the atomic chemical potentials as discussed in the current work. Finally, Stoyanova et al.\cite{Stoyanova2010} reported that in LiCo$_{1-2x}$Ni$_x$Mn$_x$O$_2$ with $x<0.05$, which can be regarded as (Ni,Mn)-doped \ce{LiCoO2}, the dopants are stable as Ni$^{3+}$ and Mn$^{4+}$, whereas in highly doped samples, $x=0.10$, they are stable as Ni$^{2+}$ and Mn$^{4+}$. This can be understood as the following: At low dopant concentration, the probability of Ni and Mn being in the proximity of each other is low; as a result, the dopants are predominantly isolated defects, i.e., stable as Ni$_{\rm Co}^0$ (i.e., Ni$^{3+}$) and Mn$_{\rm Co}^+$ (i.e., Mn$^{4+}$) as discussed earlier. At higher concentrations, the observation can be understood in terms of our results for the (Ni,Mn) pair in which the dopants are stable as Ni$^{2+}$ and Mn$^{4+}$.

\subsection{\ce{LiNiO2}}

\begin{figure}
\centering
%\vspace{-1cm}
%\hspace{-.5cm}
\includegraphics[width=8.6cm]{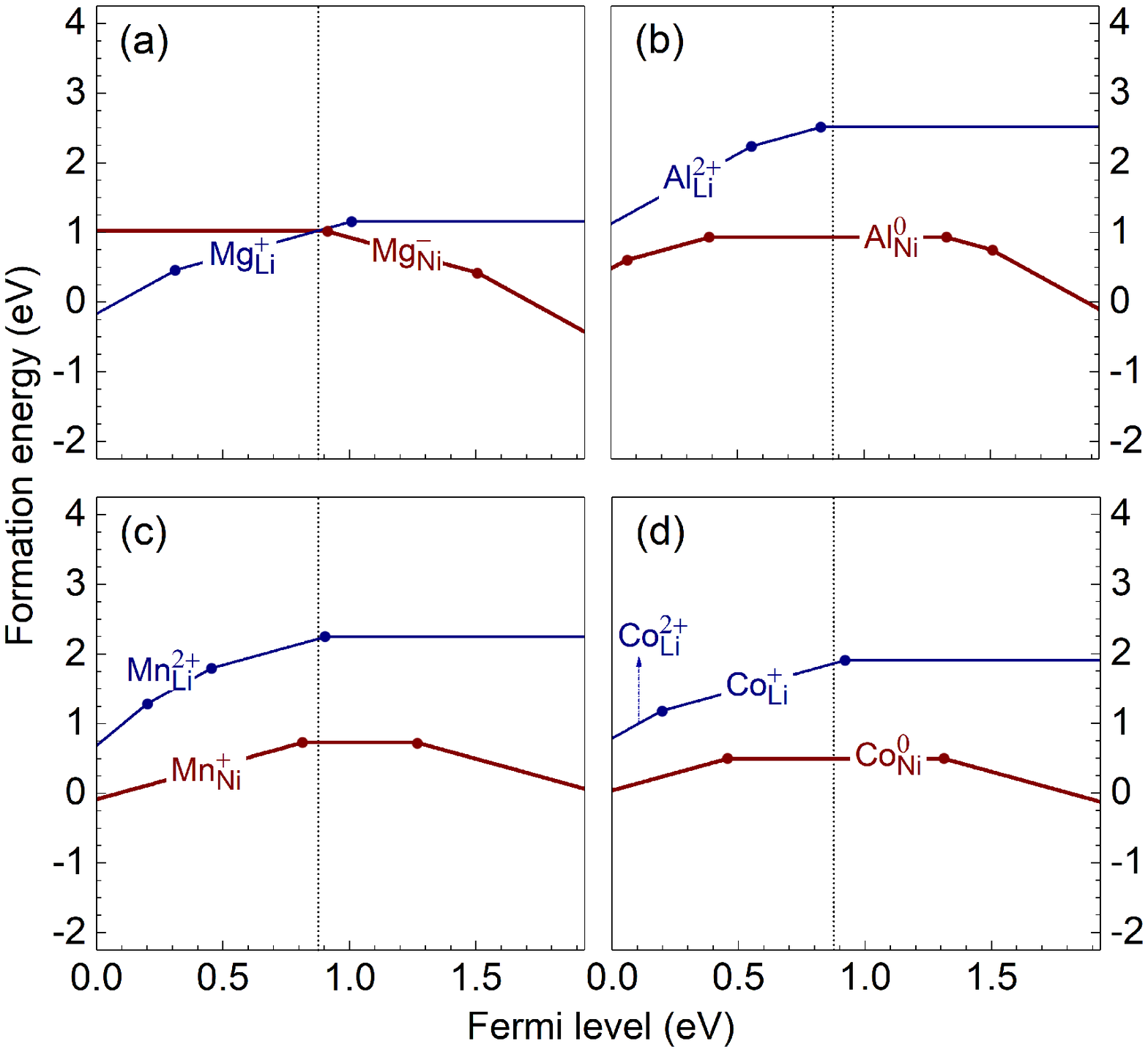}
%\vspace{0cm}
\caption{Formation energies of substitutional impurities at the Li and Ni lattice sites in \ce{LiNiO2} obtained at point $X$ [marked by a cross in the chemical-potential diagram in Fig.~\ref{fig;chempot}(b)], plotted as a function of Fermi level from the VBM to the CBM of the undoped compound: (a) Mg, (b) Al, (c) Mn, and (d) Co. The slope in the energy plots indicates the charge state ($q$). For each defect, only the true charge states are indicated. The vertical dotted line marks the Fermi level of undoped \ce{LiNiO2}, $\mu_{e}^{\rm{int}}$, determined by the intrinsic point defects.~\cite{Hoang2014JMCA}}
\label{fig;lno}
\end{figure}

Figure~\ref{fig;lno} shows the formation energies of substitutional Mg, Al, Mn, Fe, and Co impurities in \ce{LiNiO2}, obtained under conditions at point $X$ in Fig.~\ref{fig;chempot}(b). We find that each impurity has only one true charge state, as indicated in the figure, except Co$_{\rm Li}$ which can be stable as Co$_{\rm Li}^+$ (i.e., Co$^{2+}$ at the Li site) or Co$_{\rm Li}^{2+}$ (i.e., Co$^{3+}$ at the Li site) though the $2+$ charge state is only stable in the range of the Fermi-level values far away from $\mu_{e}^{\rm{int}}$, the Fermi level of undoped \ce{LiNiO2},\cite{Hoang2014JMCA} and thus not really relevant. Other (nominal) charge states are defect complexes consisting of the elementary defects and hole ($\eta^+$) or electron ($\eta^-$) polarons. Note that, in \ce{LiNiO2}, Ni is stable as low-spin Ni$^{3+}$ and the polaron $\eta^+$ ($\eta^-$) corresponds to low-spin Ni$^{4+}$ (Ni$^{2+}$) at the Ni site.\cite{Hoang2014JMCA} Mn$_{\rm Ni}^+$ (i.e., Mn$^{4+}$ at the Ni site), for example, is an elementary defect, whereas Mn$_{\rm Ni}^0$ is a complex of Mn$_{\rm Ni}^+$ and $\eta^-$; Co$_{\rm Ni}^0$ (i.e., low-spin Co$^{3+}$ at the Ni site) is an elementary defect, Co$_{\rm Ni}^+$ (Co$_{\rm Ni}^-$) is a complex of Co$_{\rm Ni}^0$ and $\eta^+$ ($\eta^-$).

\begin{figure}
\centering
%\vspace{-1cm}
%\hspace{-.5cm}
\includegraphics[width=8.6cm]{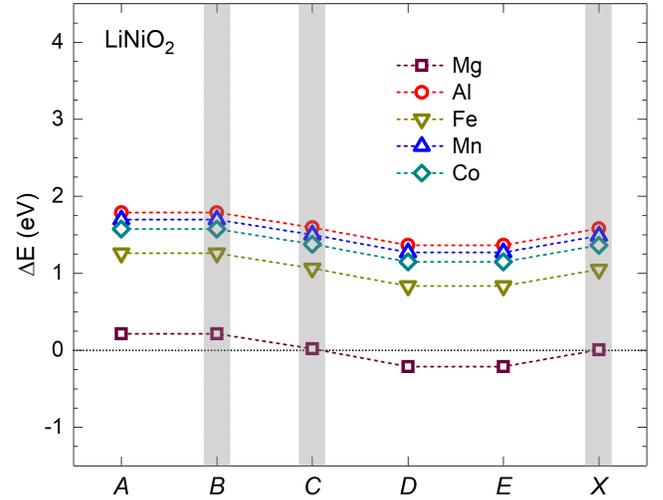}
%\vspace{0cm}
\caption{Difference between the formation energies at the Ni and Li sites, obtained under the conditions at points $A$$-$$E$ and $X$ in Fig.~\ref{fig;chempot}(b). $\Delta E > 0$ means the impurity is energetically more favorable at the Ni site. The results under more realistic synthesis conditions (points $B$, $C$, and $X$) are highlighted.}
\label{fig;lno;diff}
\end{figure}

Figure~\ref{fig;lno;diff} shows the formation-energy difference at the Fermi level $\mu_{e}^{\rm{int}}$ of undoped \ce{LiNiO2} between the Ni and Li sites. The results indicate that all the impurities, except Mg, is energetically more favorable at the Ni site. Specifically, as isolated defects, Al is most stable as Al$_{\rm Ni}^0$ (i.e., Al$^{3+}$ at the Ni site), Fe as Fe$_{\rm Ni}^0$ (i.e., high-spin Fe$^{3+}$ at the Ni site), Mn as Mn$_{\rm Ni}^+$, and Co as Co$_{\rm Ni}^0$, all independent of the atomic chemical potentials. Mg is most stable as Mg$_{\rm Ni}^-$ (i.e., Mg$^{2+}$ at the Ni site) at points $A$ and $B$, Mg$_{\rm Li}^+$ (i.e., Mg$^{2+}$ at the Li site) at points $D$ and $E$, and on both the Ni and Li sites with almost equal concentrations at $C$ and $X$, see Fig.~\ref{fig;lno;diff}. Interestingly, we find that the $\Delta E$ curves for the impurities follow the same trend and are different from one another by only a constant. This is due to the fact that (i) $\mu_{e}^{\rm{int}}$ of undoped \ce{LiNiO2} is always determined by $\eta^+$ and $\eta^-$ whose formation energies are independent of the atomic chemical potentials, i.e., $\mu_{e}^{\rm{int}}$ is a constant,\cite{Hoang2014JMCA} and (ii) the charge-state difference between the most stable configuration at the Ni site and that at the Li site is also a constant. As a result, $\Delta E$ for the impurities in \ce{LiNiO2} depends on the atomic chemical potentials only through the term $\mu_{\rm Li} - \mu_{\rm Ni}$ which varies with different points in the chemical-potential diagram in Fig.~\ref{fig;chempot}(b) but is independent of the dopants' identity.

Explicit calculations of a neutral (Co,Mn) pair show that the impurities are stable as (Co$^{3+}$,Mn$^{4+}$) in \ce{LiNiO2}. Besides, they turn one Ni$^{3+}$ of the host compound into Ni$^{2+}$; i.e., there is charge transfer between Mn and one of the Ni host atoms. The whole (Co$_{\rm Ni}$,Mn$_{\rm Ni}$)$^0$ complex can be regarded as consisting of Co$_{\rm Ni}^{0}$, Mn$_{\rm Ni}^+$, and $\eta^-$. This is consistent with the result discussed above that, as isolated defects, Co and Mn are most stable as Co$_{\rm Ni}^{0}$ and Mn$_{\rm Ni}^+$, respectively, and the fact that $\eta^-$ is easy to form in \ce{LiNiO2}.\cite{Hoang2014JMCA} The lowest-energy configuration of the complex corresponds to the closest distance (2.89 {\AA}) between the oppositely charged Mn$_{\rm Ni}^+$ and $\eta^-$. In this configuration, the complex has a binding energy of 0.70 eV with respect to its isolated constituents. For comparison, a neutral (Co,Al) pair is found to be stable as (Co$^{3+}$,Al$^{3+}$), i.e., Co$_{\rm Ni}^0$ and Al$_{\rm Ni}^{0}$ defects. The total energy of the pair is almost independent of the pair distance ($E_b \sim 0$ eV).

Like in the case of \ce{LiCoO2}, the impurities in \ce{LiNiO2} may occur as complexes with the intrinsic point defects. Explicit calculations for all possible low-energy neutral complexes between the impurity and relevant intrinsic defects are carried out for \ce{LiNiO2} and the results are summarized in Table~\ref{tab;models}. We find that under more realistic synthesis conditions, such as in the region enclosing approximately points $B$, $C$, and $X$ in Fig.~\ref{fig;chempot}(b), Mg can be present in the material in the form of the neutral complex Mg$_{\rm Ni}^-$$-$$\eta^+$ ($E_b = 0.54$ eV) and Mn in the form of Mn$_{\rm Ni}^+$$-$$\eta^-$ ($E_b = 0.57$ eV). These two impurities are thus incorporated as negatively (positively) charged defects that are charge-compensated by hole (electron) polarons. The presence of the polarons as accompanying intrinsic defects of the impurities under all synthesis conditions is consistent with our previous study showing that $\eta^+$ and $\eta^-$ are the lowest-energy intrinsic defects in \ce{LiNiO2}, a property that originates from the ability of low-spin Ni$^{3+}$ in the layered oxide to undergo charge disproportionation: 2Ni$^{3+}$ $\rightarrow$ Ni$^{4+}$ + Ni$^{2+}$.\cite{Hoang2014JMCA} The other impurities can be present as Al$_{\rm Ni}^0$, Fe$_{\rm Ni}^0$, and Co$_{\rm Ni}^0$, respectively; i.e., they are trivalent impurities; see Table~\ref{tab;models}. Other defect complexes listed in Table~\ref{tab;models} include Mg$_{\rm Li}^+$$-$$\eta^-$ ($E_b = 0.38$ eV), Mg$_{\rm Ni}^-$$-$Ni$_{\rm Li}^+$ ($E_b = 0.44$ eV), and Mn$_{\rm Ni}^+$$-$Li$_{\rm Ni}^{2-}$$-$$\eta^+$ ($E_b = 1.62$ eV). The electronic structure of selected heavily doped \ce{LiNiO2} systems is reported in Ref.~\citenum{SM}.   

Experimentally, Pouillerie et al.~\cite{Pouillerie2000} reported that in Mg-doped \ce{LiNiO2} samples, LiNi$_{1-y}$Mg$_y$O$_2$, a certain amount of Mg goes into the Li site, especially at $y\geq 0.10$. This is consistent with our results showing that Mg can be at the Ni and/or Li sites, see Fig.~\ref{fig;lno;diff} and Table~\ref{tab;models}. Other impurities were found to be incorporated at the Ni site,\cite{Delmas1992SSI,Mukai2010JSSC} again, consistent with the computational results.

\subsection{\ce{LiMnO2}}

\begin{figure}
\centering
%\vspace{-1cm}
%\hspace{-.5cm}
\includegraphics[width=8.6cm]{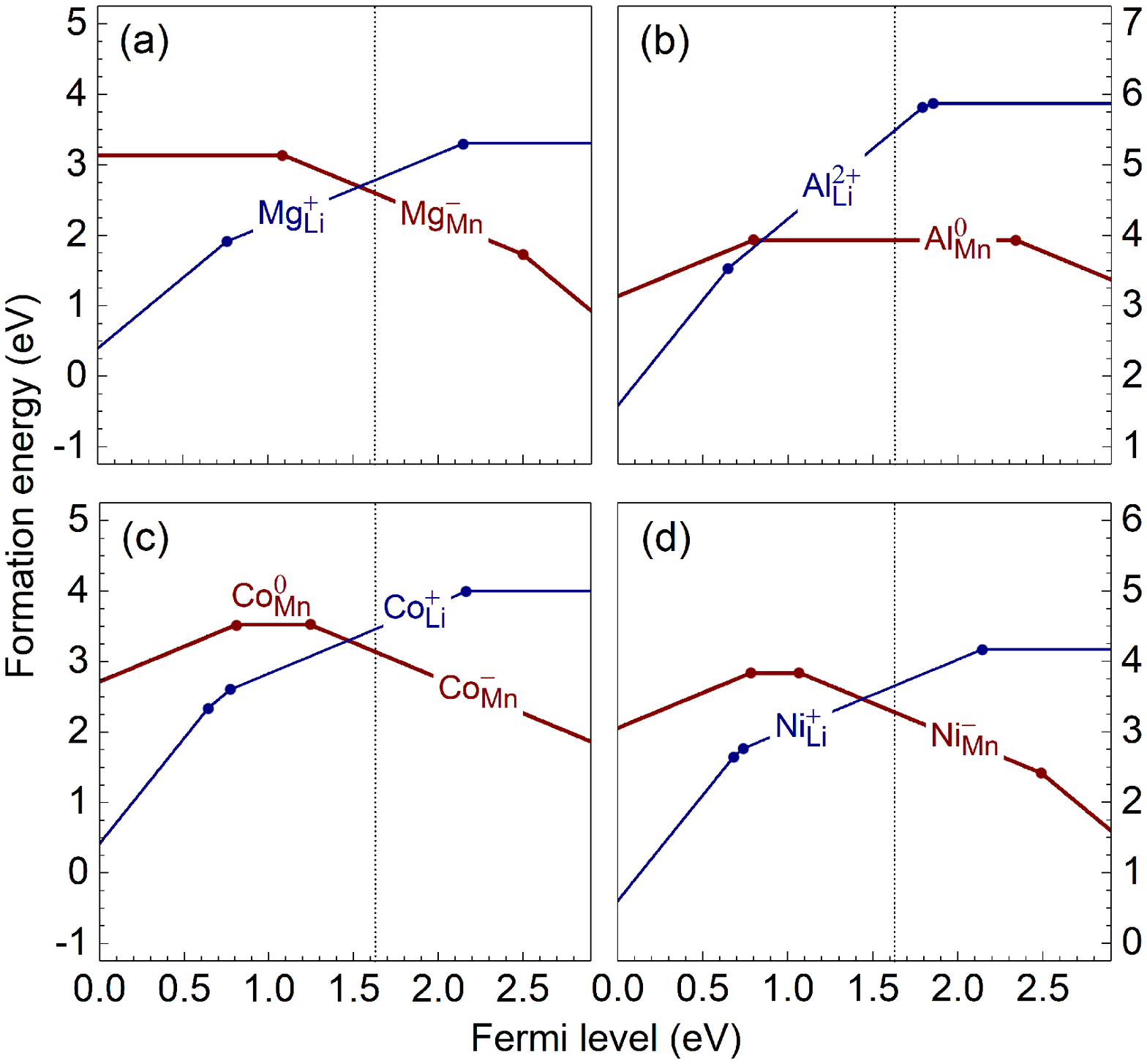}
%\vspace{0cm}
\caption{Formation energies of substitutional impurities at the Li and Mn sites in \ce{LiMnO2} obtained at point $X$ [marked by a cross in the chemical-potential diagram in Fig.~\ref{fig;chempot}(c)], plotted as a function of Fermi level from the VBM to the CBM of the undoped compound: (a) Mg, (b) Al, (c) Co, and (d) Ni. The slope in the energy plots indicates the charge state ($q$). For each defect, only the stable charge states are indicated. The vertical dotted line marks the Fermi level of undoped \ce{LiMnO2}, $\mu_{e}^{\rm{int}}$, determined by the intrinsic point defects.\cite{Hoang2015PRA}}
\label{fig;lmo}
\end{figure}

Figure~\ref{fig;lmo} shows the formation energies of substitutional Mg, Al, Fe, Co, and Ni impurities at the Mn and Li sites in \ce{LiMnO2}, obtained under the conditions at point $X$ in Fig.~\ref{fig;chempot}(c); defect configurations with true charge states, i.e., the elementary defects, are indicated. Other charge states are complexes of the elementary defects and hole ($\eta^+$) or electron ($\eta^-$) polarons. Note that, Mn is stable as high-spin Mn$^{3+}$ in \ce{LiMnO2} and the polaron $\eta^+$ ($\eta^-$) corresponds to Mn$^{4+}$ (high-spin Mn$^{2+}$) at the Mn site.\cite{Hoang2015PRA}

\begin{figure}
\centering
%\vspace{-1cm}
%\hspace{-.5cm}
\includegraphics[width=8.6cm]{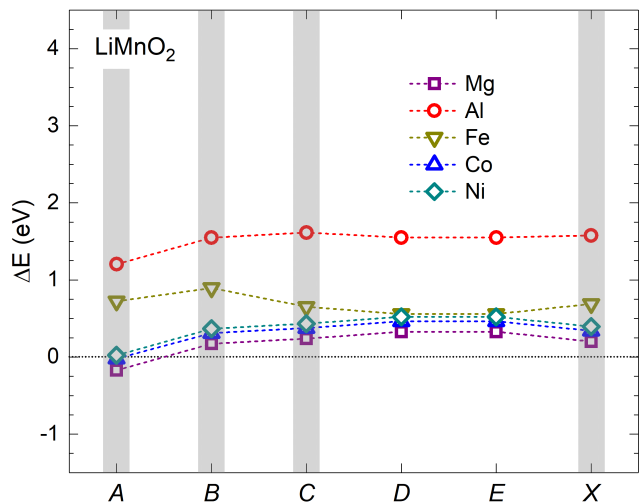}
%\vspace{0cm}
\caption{Difference between the formation energies at the Mn and Li sites, obtained under the conditions at points $A$$-$$E$ and $X$ in Fig.~\ref{fig;chempot}(c). $\Delta E > 0$ means the impurity is energetically more favorable at the Mn site. The results obtained under more realistic synthesis conditions are highlighted.}
\label{fig;lmo;diff}
\end{figure}

Figure~\ref{fig;lmo;diff} shows the formation-energy difference at the Fermi level $\mu_{e}^{\rm{int}}$ of undoped \ce{LiMnO2} between the Mn and Li sites. The impurities are found to be more favorable at the Mn site, except under the conditions at point $A$ where Mg at the Li site is slightly more favorable and Co and Ni can be on both the lattice sites. Specifically, as isolated defects, Al is stable as Al$_{\rm Mn}^0$ (i.e., Al$^{3+}$ at the Mn site), Fe as Fe$_{\rm Mn}^0$ (i.e., high-spin Fe$^{3+}$ at the Mn site), and Ni as Ni$_{\rm Mn}^-$ (i.e., high-spin Ni$^{2+}$ at the Mn site). Mg is found to be most stable as Mg$_{\rm Mn}^-$ (i.e., Mg$^{2+}$ at the Mn site) at points $B-E$ and $X$ or Mg$_{\rm Li}^+$ (i.e., Mg$^{2+}$ at the Li site) at points $A$. Finally, Co is most stable as Co$_{\rm Mn}^-$ (i.e., high-spin Co$^{2+}$ at the Mn site) at points $B-E$ and $X$ or Co$_{\rm Li}^+$ (i.e., high-spin Co$^{2+}$ at the Li site) at point $A$. Ni and Co are thus stable as Ni$^{2+}$ and Co$^{2+}$, respectively, independent of the atomic chemical potentials.

For comparison, Prasad et al.,\cite{Prasad2005} based on an analysis of the metal-oxygen bond lengths, also found that Ni and Co are stable as Ni$^{2+}$ and Co$^{2+}$ in layered \ce{LiMnO2}. Regarding the lattice site preference, Kong et al.\cite{Kong2015JPCC} showed that Ni and Co are energetically more favorable at the Mn site, which is in general consistent with our results (except those obtained under the conditions at point $A$). They, however, also indicated that Ni has a slightly larger tendency to occupy the Li site compared to Co, which appears to contradict our results reported in Fig.~\ref{fig;lmo;diff} where $\Delta E$ associated with Ni is slightly higher than that with Co. The discrepancy is likely due to the different computational approaches adopted in the two studies. 

Explicit calculations of a neutral (Ni,Co) pair on the Mn sublattice shows that the impurities are stable as Ni$^{2+}$ and Co$^{3+}$ and one of the Mn$^{3+}$ ions in the host compound is turned into Mn$^{4+}$. There is thus charge transfer from Ni to one of the Mn host atoms. The whole pair is, in fact, a complex of Ni$_{\rm Mn}^{-}$, Co$_{\rm Mn}^{0}$, and $\eta^+$. Though Co is most stable as Co$^{2+}$ as an isolated defect as discussed above, here it is energetically more favorable as Co$^{3+}$ due to the presence of Ni$^{2+}$ and the formation of $\eta^+$ (i.e., Mn$^{4+}$). The most stable configuration of the defect complex corresponds to the shortest distance (2.81 {\AA}) between the oppositely charged defects Ni$_{\rm Mn}^{-}$ and $\eta^+$. In this configuration, the complex has a binding energy of 0.37 eV with respect to its isolated constituents. 

We also carry out calculations for all possible low-energy neutral complexes between the impurity and relevant intrinsic defects in \ce{LiMnO2}. Table~\ref{tab;models} summarizes the lowest-energy defect models for the doped materials. Under more realistic conditions, such as in the region enclosing approximately points $A$, $B$, $C$, and $X$ in Fig.~\ref{fig;chempot}(c), Mg can be present in the material in the form of the neutral complex Mg$_{\rm Mn}^-$$-$$\eta^+$ ($E_b = 0.35$ eV), Mg$_{\rm Li}^+$$-$$V_{\rm Li}^-$ ($E_b = 0.29$ eV), or Mg$_{\rm Mn}^-$$-$Mn$_{\rm Li}^+$ ($E_b = 0.30$ eV), depending on the specific conditions. Al is most stable as as Al$_{\rm Mn}^0$, Fe as Fe$_{\rm Mn}^0$, and Co as Co$_{\rm Mn}^0$ or Co$_{\rm Mn}^-$$-$Mn$_{\rm Li}^+$ ($E_b = 0.28$ eV). Finally, Ni can be present in the material as Ni$_{\rm Mn}^-$$-$$\eta^+$ ($E_b = 0.33$ eV) or Ni$_{\rm Mn}^-$$-$Mn$_{\rm Li}^+$ ($E_b = 0.31$ eV). It should be noted that, in addition to these lowest-energy defect models, those that have slightly higher energies are also reported in the footnotes of Table~\ref{tab;models}. The other defect complexes associated with \ce{LiMnO2} listed in Table~\ref{tab;models} include Mg$_{\rm Li}^+$$-$Li$_{\rm Mn}^{2-}$$-$$\eta^+$ ($E_b = 1.06$ eV), Co$_{\rm Li}^+$$-$$V_{\rm Li}^-$ ($E_b = 0.26$ eV), Co$_{\rm Mn}^-$$-$Li$_i^+$ ($E_b = 0.31$ eV), Ni$_{\rm Li}^+$$-$$V_{\rm Li}^-$ ($E_b = 0.24$ eV), and Ni$_{\rm Mn}^-$$-$Li$_i^+$ ($E_b = 0.32$ eV). All these complexes, except Mg$_{\rm Li}^+$$-$Li$_{\rm Mn}^{2-}$$-$$\eta^+$, have rather small binding energies; as a result, they may dissociate into their isolated constituents, especially under thermal equilibrium at high temperatures.\cite{walle:3851} The electronic structure of selected heavily doped \ce{LiMnO2} systems is also calculated and reported in Ref.~\citenum{SM}.

\section{Conclusions}

We have carried out a detailed study of doping in layered oxides \ce{LiMO2}, mainly in the dilute doping limit, using first-principles defect calculations based on a hybrid DFT/Hartree-Fock approach. We find that Al, Fe, and Mn impurities are more favorable on the Co sublattice in \ce{LiCoO2}, whereas Mg and Ni can be on the Co and/or Li sublattices depending on the synthesis conditions. In \ce{LiNiO2}, Al, Fe, Mn, and Co are more favorable on the Ni sublattice; Mg can be incorporated on the Ni and/or Li sublattices. Finally, Mg, Al, Fe, Co, and Ni are energetically more favorable on the Mn sublattice in \ce{LiMnO2}, except under the synthesis conditions where the system is close to an equilibrium with \ce{Li2MnO3} and \ce{Mn3O4} where Mg, Co, and Ni can be on the Mn and/or Li sublattices.

More importantly, we find that the lattice site preference is dependent not only on the ionic-radius difference between the doping element and the substituted host ion, which is related to the charge and spin states of the dopant at the substituted lattice site, but also on the relative abundance of the host compound's constituting elements in the synthesis environment. On the basis of the structure and energetics of the impurities and their complexes with intrinsic point defects, we have determined all possible low-energy impurity-related defect complexes in the doped materials. These defect models are useful for further analyses of the doped materials and in interpreting the experimental observations. Finally, the lightly doped \ce{LiMO2} materials considered here can be regarded as model systems for understanding the more complex, mixed-metal, \ce{LiMO2}-based battery electrode materials.

\begin{acknowledgments}

This work was supported in part by the U.S.~Department of Energy Grant No.~DE-SC0001717 and made use of computing resources in the Center for Computationally Assisted Science and Technology at North Dakota State University.

\end{acknowledgments}

% Create the reference section using BibTeX:
%\bibliography{layeredoxides}
%

\end{document}